\documentclass[12pt]{article}
\usepackage{amsmath}
\usepackage{amsfonts}
\usepackage{amssymb}
\usepackage{graphicx}

\begin{document}

\title{Acceleration in Weyl integrable spacetime}
\author{John Miritzis \\
Department of Marine Sciences, University of the Aegean \\
University Hill, Mytilene 81100, Greece\\
e-mail: imyr@aegean.gr}

\date{}
\maketitle

\begin{abstract}
We study homogeneous and isotropic cosmologies in a Weyl spacetime. It is
shown that in Weyl integrable spacetime, the corresponding scalar field may
act as a phantom field. In this circumstance the Weyl field gives rise to a
late accelerated expansion of the Universe for all initial conditions and
for a wide range of the parameters.
\end{abstract}

\section{Introduction}

A major effort in theoretical cosmology is a satisfactory explanation of the
late-time acceleration of the universe. No convincing theory has yet been
constructed to explain this state of affairs. The proposed models,
characterized by a departure from conventional cosmology, either assume the
existence of dark energy \cite{sast,pera}, or require a modification of
general relativity at cosmological distance scales \cite{chib,cdtt}, (cf. 
\cite{cst}-\cite{bcno} for comprehensive reviews and references). Less
explored is the idea that the geometry of spacetime is not the so far
assumed Lorentz geometry (see for example \cite{cct,ccsv,miri}). In this
circumstance we need not only look for a new idea, but also revisit every
type of the old models from a new viewpoint. Weyl geometry is certainly one
candidate to be reviewed and reanalyzed. According to the Ehlers, Pirani and
Schild proposed set of axioms \cite{eps}, Weyl geometry is the natural
space-time structure that follows from a small number of basic assumptions
of two observable quantities: the worldlines of light rays and free falling
particles \cite{mfff,fafr}.

We recall that a Weyl space is a manifold endowed with a metric $\mathbf{g}$
and a symmetric connection $\mathbf{\nabla }$ which are interrelated via 
\begin{equation*}
\nabla _{\mu }g_{\alpha \beta }=-Q_{\mu }g_{\alpha \beta },
\end{equation*}%
where the 1-form $Q_{\mu }$ is customarily called Weyl covariant vector
field (see for example \cite{scho,gnc} and the Appendix in \cite{miri} for a
detailed exposition of the techniques involved in Weyl geometry). We denote
by $D$ the Levi-Civita connection of the metric $g_{\alpha \beta }$. If $%
Q_{\mu }$ is a gradient, i.e., if $Q_{\mu }=\partial _{\mu }\phi $ for some
scalar function $\phi ,$ the corresponding space is called Weyl integrable
spacetime. Integrable Weyl geometry does not suffer from the so-called
second clock effect and for that reason it has been used in some approaches
to gravitation and cosmology \cite{nohe}-\cite{quir}.

In this short paper we study Friedmann-Robertson-Walker (FRW) cosmologies in
a Weyl framework. We demonstrate that the simple model based on the
Lagrangian (\ref{action}), may provide a mechanism of late accelerating
expansion.

\section{A simple Lagrangian}

The field equations obtained from the Lagrangian $L=R$ in a Weyl spacetime
constitute the generalization of the Einstein equations in vacuum (cf. eqs
(30) and (31) in \cite{cmq}). More general Lagrangians have also been
suggested, especially in the context of the Palatini formalism, for example $%
L=e^{-\phi}R$ in \cite{posa}; $L=R+\beta R^{2}-2\Lambda$ gravity for FRW
models was investigated in \cite{gnd}; see also \cite{rfp} for a formulation
of a theory invariant with respect to Weyl transformations and \cite{nrm}
for the inclusion of torsion. There is however an alternative view, namely
that the pair $\left( Q,\mathbf{g}\right) $ which defines the Weyl spacetime
also enters into the gravitational theory and therefore, the field $Q$ must
be contained in the Lagrangian independently from $\mathbf{g.}$ In the case
of integrable Weyl geometry, i.e. when $Q_{\mu}=\partial_{\mu}\phi$ where $%
\phi$ is a scalar field, the pair $\left( \phi,g_{\mu\nu}\right) $
constitutes the set of fundamental geometrical variables. We stress that the
nature of the scalar field is purely geometric.

A simple Lagrangian involving the set $\left( \phi ,g_{\mu \nu }\right) $ is
given by 
\begin{equation}
L=R+\xi \nabla _{\mu }Q^{\mu }+L_{m},  \label{action}
\end{equation}%
where $\xi $ is a constant and $L_{m}$ corresponds to the Lagrangian
yielding the energy-momentum tensor of a perfect fluid. Motivations for
considering theory (\ref{action}) can be found in \cite{nove,sasa} (see also 
\cite{kome,agro1,agro} for a multidimensional approach and \cite{oss} for an
extension of (\ref{action}) to include an exponential potential function of $%
\phi $). The inclusion of the term $\xi \nabla _{\mu }Q^{\mu }$ in the
Lagrangian is not an arbitrary choice, as is explained in \cite{nobe}. In
fact, apart from the scalar curvature $R,$ there are two invariants with
dimension of inverse squared length that can be constructed with the
fundamental geometrical variables of a Weyl space: $\nabla _{\mu }Q^{\mu }$
and $Q_{\mu }Q^{\mu }.$ Now the derivative operator $\nabla _{\mu }$ can be
expressed in terms of the Levi-Civita derivative operator $D_{\mu },$ so
that, $\nabla _{\mu }Q^{\mu }=D_{\mu }Q^{\mu }+2Q_{\mu }Q^{\mu },$ (see the
list of identities in the Appendix in \cite{miri}). Therefore the two
invariants reduce to one. We conclude that the simplest vacuum Lagrangian
constructed from purely geometric quantities has the form $L=R+\xi \nabla
_{\mu }Q^{\mu }$.

By varying the action corresponding to (\ref{action}) with respect to both $%
g_{\mu \nu }$ and $\phi $ one obtains \cite{sasa,oss} 
\begin{equation}
\overset{\circ }{G}_{\mu \nu }=\frac{3-4\xi }{2}\left( \partial _{\mu }\phi
\partial _{\nu }\phi -\frac{1}{2}\left( \partial _{\alpha }\phi \partial
^{\alpha }\phi \right) g_{\mu \nu }\right) +T_{\mu \nu },  \label{nove}
\end{equation}%
and%
\begin{equation}
\overset{\circ }{\square }\phi =\frac{1}{4\xi -3}\rho ,  \label{emsf}
\end{equation}%
where the accent $\overset{\circ }{}$ denotes a quantity formed with the
Levi-Civita connection, $D$. As mentioned above, ordinary matter described
by $T_{\mu \nu }$ is a perfect fluid with energy density $\rho $ and
pressure $p$. Setting 
\begin{equation*}
\lambda =\frac{4\xi -3}{2},
\end{equation*}%
we note that for $\lambda <0$ the field equations are formally equivalent to
general relativity with a massless scalar field coupled to a perfect fluid%
\footnote{%
The case $\lambda =-3/2,$ i.e., the simple theory $L=R+L_{m},$ was analyzed
in \cite{miri} under the severe assumption of separate conservation of the
two fluids. It was shown that for all FRW models, the Weyl fluid has a
significant contribution only near the cosmological singularities. In
expanding models the \textquotedblleft real\textquotedblright\ fluid always
dominates at late times and therefore the contribution of the Weyl fluid to
the total energy-momentum tensor is important only at early times. Similar
results were obtained in \cite{miri1}, under the assumption of energy
exchange between the two fluids.}. On the other hand the case $\lambda >0$,
is more interesting, because the Weyl field $\phi $ plays the role of a
phantom scalar field characterized by the \textquotedblleft
wrong\textquotedblright\ sign of the kinetic term. We emphasize that the
scalar field has a geometric nature in Weyl spacetime and therefore, no
restriction exists for the sign of the value of $\lambda $. In the following
we consider only the case of positive $\lambda $.

\section{Acceleration of FRW models}

We assume an initially expanding FRW universe with expansion scale factor $%
a\left( t\right) $ and Hubble function $H=\dot{a}/a$. We adopt the metric
and curvature conventions of \cite{wael}. An overdot denotes differentiation
with respect to time $t,$ and units have been chosen so that $c=1=8\pi G.$
Ordinary matter is described by a perfect fluid supplemented with an
equation of state,%
\begin{equation*}
p=(\gamma-1)\rho,\ \ \ 0<\gamma<2.
\end{equation*}
The field equations (\ref{nove}) and (\ref{emsf}) imply the Friedmann
equation,%
\begin{equation}
H^{2}+\frac{k}{a^{2}}=\frac{1}{3}\rho-\frac{\lambda}{6}\dot{\phi}^{2},
\label{fri1jm}
\end{equation}
with $k=-1,0,+1,$ the Raychaudhuri equation, 
\begin{equation*}
\dot{H}=-\frac{\gamma}{2}\rho+\frac{\lambda}{2}\dot{\phi}^{2}+\frac{k}{a^{2}}%
,
\end{equation*}
the equation of motion of the scalar field, 
\begin{equation*}
\ddot{\phi}+3H\dot{\phi}=-\frac{1}{2\lambda}\rho,
\end{equation*}
and the conservation equation, 
\begin{equation*}
\dot{\rho}=-3\gamma\rho H-\frac{1}{2}\rho\dot{\phi}.
\end{equation*}

We introduce expansion-normalized variables 
\begin{equation}
x=\frac{\dot{\phi}}{\sqrt{6}H},\ \ \ \ \Omega=\frac{\rho}{3H^{2}},\ \ \ \ K=%
\frac{k}{H^{2}a^{2}},  \label{expa}
\end{equation}
and a new time variable $\tau$ defined by $\tau=\ln a.$ For flat, $k=0,$
models the dynamical system becomes 
\begin{align}
x^{\prime} & =-3x-\sqrt{\frac{3}{2}}\frac{1}{2\lambda}\Omega+x\left( \frac{%
3\gamma}{2}\Omega-3\lambda x^{2}\right) ,  \notag \\
\Omega^{\prime} & =\Omega\left( -3\gamma-\sqrt{\frac{3}{2}}x+3\gamma
\Omega-6\lambda x^{2}\right) ,  \label{sys2}
\end{align}
where a prime denotes differentiation with respect to $\tau$. The evolution
of the Hubble function is described by the equation,%
\begin{equation*}
H^{\prime}=-H\left( \frac{3\gamma}{2}\Omega-3\lambda x^{2}\right) ,
\end{equation*}
which decouples from the rest of the evolution equations (\ref{sys2}). This
is one of the merits of the introduction of the variables (\ref{expa}),
namely that it allows for the reduction of the dimension of the dynamical
system by one.

The constraint (\ref{fri1jm}) takes the form 
\begin{equation}
\Omega =1+\lambda x^{2}.  \label{const}
\end{equation}%
The last equation implies that the phase space of the dynamical system
consists of a parabola in the $x-\Omega $ plane, i.e., it is
one-dimensional. In fact, substituting (\ref{const}) into the first of (\ref%
{sys2}), we obtain the one-dimensional dynamical system 
\begin{equation}
x^{\prime }=-\sqrt{\frac{3}{2}}\frac{1}{2\lambda }+3\left( \frac{\gamma }{2}%
-1\right) x-\frac{1}{2}\sqrt{\frac{3}{2}}x^{2}+3\left( \frac{\gamma }{2}%
-1\right) \lambda x^{3}.  \label{sys1d}
\end{equation}%
The only real equilibrium of (\ref{sys1d}) is 
\begin{equation}
x_{\ast }=-\frac{1}{\sqrt{6}\left( 2-\gamma \right) \lambda }.  \label{equi}
\end{equation}%
This solution was also found by Oliveira et al \cite{oss}, in the context of
inflation. It corresponds to a \textquotedblleft scaling" solution where the
energy density of the Weyl field is a constant fraction of the total energy
density. A plot of the polynomial right-hand side of (\ref{sys1d}) shows
that the equilibrium solution (\ref{equi}) is \emph{globally asymptotically
stable}, i.e., $x_{\ast }$ \emph{is the future attractor of all solutions of
(\ref{sys1d})}.

\begin{figure}[h]
\begin{center}
\includegraphics{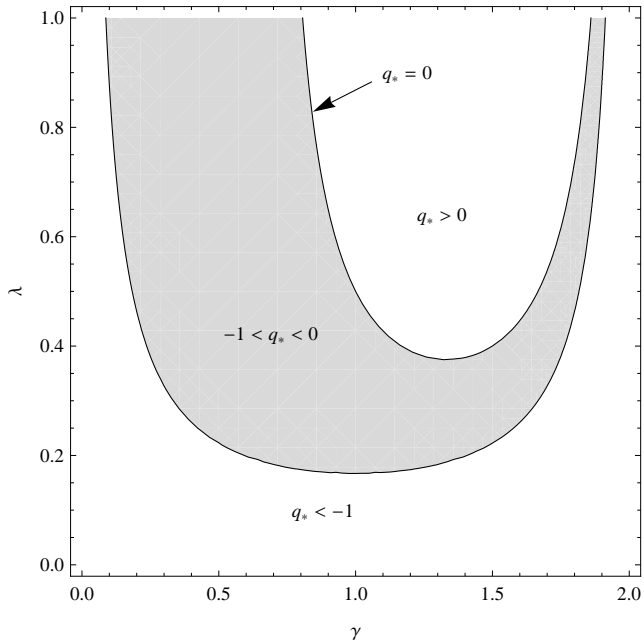}
\end{center}
\caption{In region below the parabolic curve $q_{\ast}=0$, all models
asymptotically expand with acceleration. The dark region corresponds to
models entering a phase of late accelerating expansion and avoid the big rip
singularity.}
\label{fig1jm}
\end{figure}

The decceleration parameter $q=-\ddot{a}a/\dot{a}^{2}$ at the equilibrium is
given by 
\begin{equation}
q_{\ast}=\frac{1+2\lambda\left( \gamma-2\right) \left( 3\gamma-2\right) }{%
4\lambda\left( \gamma-2\right) }.  \label{q}
\end{equation}
In Figure 1 the curve corresponding to $q_{\ast}=0,$ separates the parameter
space into two disjoint regions. The region below the parabolic curve, $%
q_{\ast}=0$, contains the values of the parameters corresponding to
accelerating expansion ($\ddot{a}>0$). It is evident that the simple model (%
\ref{action}) allows for a late accelerated expansion of the universe for a
wide range of the parameters.

\section{Final remarks}

In models with minimally coupled phantom fields, the density of the dark
energy increases with increasing scale factor and, both the scale factor and
the phantom energy density can become infinite at a finite $t$, a condition
known as the \textquotedblleft big rip\textquotedblright ,\ \cite%
{not,seva,flnos}. Since the scale factor corresponding to the asymptotic
solution (\ref{equi}) evolves as $t^{1/\left( 1+q_{\ast }\right) },$ flat
FRW models in (\ref{action}) theory may avoid the big rip singularity
provided that $-1<q_{\ast }$. We conclude that for any couple $\left( \gamma
,\lambda \right) $ in (\ref{q}) satisfying $-1<q_{\ast }<0,$ all initially
expanding universes eventually enter a phase of accelerating expansion and
avoid the big rip singularity. Inclusion of curvature, $k=\pm 1,$ increases
the dimension of the dynamical system by one. However, it can be shown that
the resulting models share the main feature of the flat model, namely that
the state of the system asymptotically approaches a stable equilibrium
corresponding to accelerating expansion. The results of this study are
purely qualitative and therefore, further investigation is necessary to
answer the question if the (\ref{action}) theory may model the whole history
of the Universe.

\section*{Acknowledgements}

I thank N. Spyrou and S. Cotsakis for useful comments during the preparation
of this work.

\end{document}